\documentclass[%
aip,
jap,
 amsmath,amssymb,
 reprint,%
]{revtex4-2}

\usepackage{graphicx}
\usepackage{dcolumn}
\usepackage{bm}
\usepackage{siunitx}
\let\siold\si
\renewcommand{\si}[1]{\,\siold{#1}}
\usepackage{mhchem}

\usepackage[utf8]{inputenc}
\usepackage[T1]{fontenc}
\usepackage{mathptmx}

\begin{document}

\preprint{AIP/123-QED}

\title[Using silicon-vacancy centers in diamond to probe the full strain
tensor]{Using silicon-vacancy centers in diamond to probe the full strain
  tensor}

\author{Kelsey M. Bates}
\affiliation{Department of Physics, University of Michigan, Ann Arbor, MI 48109,
  USA}
\author{Matthew W. Day}
\affiliation{Department of Physics, University of Michigan, Ann Arbor, MI 48109,
  USA}
\author{Christopher L. Smallwood}
\affiliation{Department of Physics, University of Michigan, Ann Arbor, MI 48109,
  USA}
\affiliation{Department of Physics and Astronomy, San Jos\'{e} State University,
  San Jos\'{e}, CA 95192, USA}
\author{Rachel C. Owen}
\affiliation{Department of Physics, University of Michigan, Ann Arbor, MI 48109,
  USA}
\author{Ronald Ulbricht}
\affiliation{Department of Molecular Spectroscopy, Max Planck Institute for
  Polymer Research, Mainz 55128, Germany}
\author{Tim Schr\"oder}
\affiliation{Department of Electrical Engineering and Computer Science, Massachusetts Institute of Technology, Cambridge, MA 02138, USA}
\affiliation{Department of Physics, Humboldt-Universit\"at zu Berlin, Newtonstrasse 15, 12489 Berlin, Germany}
\author{Edward Bielejec}
\affiliation{Sandia National Laboratories, Albuquerque, NM 87185, USA}
\author{Steven T. Cundiff}
\email[Corresponding author, e-mail:]{cundiff@umich.edu}
\affiliation{Department of Physics, University of Michigan, Ann Arbor, MI 48109,
  USA}

\date{\today}

\begin{abstract}
An ensemble of silicon vacancy centers in diamond (\ce{SiV-}) is probed using
two coherent spectroscopy techniques. Two main distinct families of \ce{SiV-}
centers are identified using multidimensional coherent spectroscopy, and these
families are paired with two orientation groups by comparing spectra from
different linear polarizations of the incident laser. By tracking the peak
centers in the measured spectra, the full diamond strain tensor is calculated
local to the laser spot. Such measurements are made at multiple points on the
sample surface and variations in the strain tensor are observed.
\end{abstract}

\maketitle

\section{\label{sec:intro}Introduction}

Color centers in diamond and other crystalline host materials have shown potential in recent years in applications as diverse as remote magnetic field sensing and imaging \cite{Taylor2008,Jensen2014,wolf2015,Ku2020}, as room-temperature qubits \cite{Ivady2017}, as single photon sources \cite{kurtsiefer2000,beveratos2001,Rogers2014,Sipahigil2014}, and even as candidates for use in novel particle detection schemes in dark matter direct detection experiments \cite{Budnik2018,rajendran2017}. One  such color center is the negatively charged silicon-vacancy center (\ce{SiV-}) in diamond. Composed of an interstitial silicon atom located between two carbon vacancies in the diamond lattice,  \ce{SiV-} centers are, to a large extent, protected from first-order perturbations by the similarity of their ground and excited states \cite{Udvarhelyi2019}. As a result, their electrostatic dipole moment vanishes, which reduces the \ce{SiV-} sensitivity to first-order perturbations. In addition, \ce{SiV-} centers exhibit weak electron-phonon coupling \cite{gorokhovsky1995,neu2011,rose2018, jahnke2015} relative to other types of color centers in diamond. These properties combine to increase dephasing times of the $1.68\si{\electronvolt}$ zero phonon line (ZPL) transition manifold to greater than $20\si{\nano\s}$ in some cases  \cite{Becker2018}.  Due to the robustness and isolation which the diamond lattice provides, the \ce{SiV-} is a promising candidate for use in many quantum information technology, remote sensing, and photonics applications.

In addition to these applications, it has recently been demonstrated that color
centers in diamond can be used to sense strain in their environments
\cite{Cai2014, Trusheim2016, Meesala2018} by carefully tracking the frequencies
of various optical or magnetic resonances as a function of sample position, and
then mapping that data to strain. This sensitivity is potentially useful in applications where remotely measuring the deviation of structural components from expected stresses and strains is desired. For example, due to the ubiquity of diamond anvil cells as devices to apply strain to study a wide variety of material samples \cite{Jayaraman1983}, having direct color center strain imaging probes \textit{in situ} would represent a tremendous advantage over traditional strain estimation techniques when studying small volume samples. 

To this end, we demonstrate a method employing silicon-vacancy centers in
diamond to spectroscopically map all strain tensor elements. We employ two
different coherent spectroscopic techniques to track the influence of strain on
the \ce{SiV-} zero phonon line transitions in our sample. By combining linear
spectra with nonlinear, multi-dimensional spectra
\cite{Cundiff2013,Smallwood2018} as a function of polarization, we are able to
provide an estimate of the ensemble averaged strain in the presence of
inhomogenaiety in our sample. Our methods are widely applicable and easily
replicable with the recent advent of commercially available coherent
spectrometers.

\section{\label{sec:method}Experimental Methods}

\subsection{\label{sec:spectroscopy}Spectroscopic Techniques}

\begin{figure}[h!]\centering\includegraphics{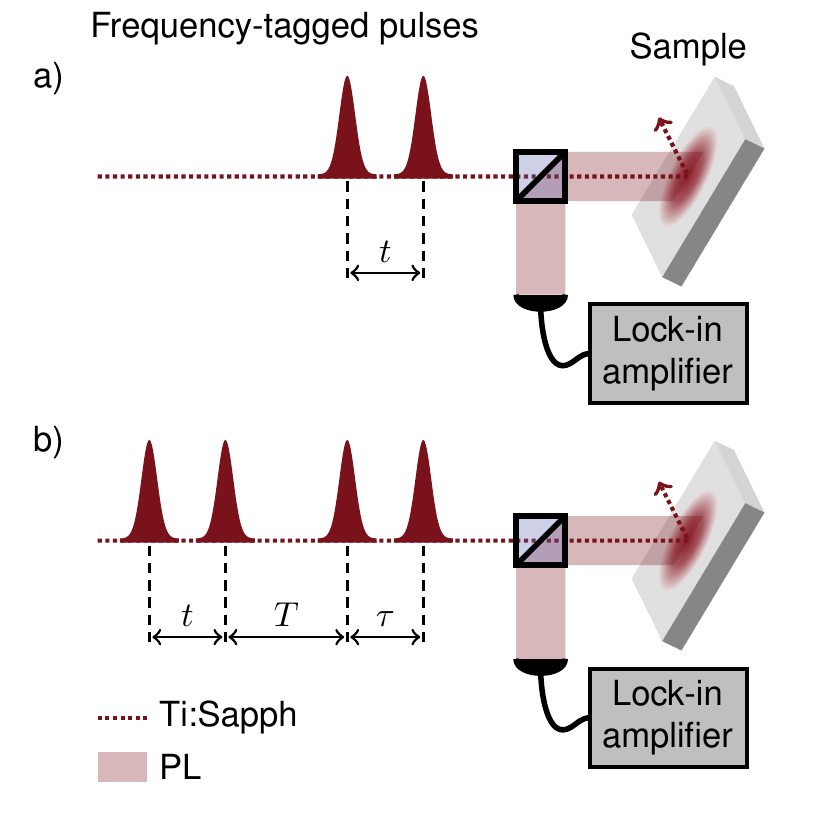}
  \caption{\label{fig:MDCS} A depiction of the experimental setup, including
    both (a) two-pulse correlation (TPC) measurements and (b) multidimensional
    coherent spectroscopy (MDCS) measurements.} 
\end{figure}

We study an ensemble of \ce{SiV-} centers with two different coherent
spectroscopy techniques. The first is a two-pulse correlation
(TPC) measurement. Here, two collinear optical pulses from a Ti:Sapph laser,
which are frequency tagged with acousto-optic modulators in a Mach-Zhender
interferometer, are sent to the sample, as depicted in Fig. \ref{fig:MDCS}
(a). Using a lock-in amplifier, we measure the modulated photoluminescence from
the sample as a function of the temporal separation ($t$) between the
pulses. Additionally, a continuous wave laser is co-propagated with the Ti:Sapph
pulses to act as a phase reference for the lock-in amplifier and to remove the
effects of mechanical fluctuations in the experiment. The photoluminescence
intensity is Fourier transformed with respect to $t$, to yield a
one-dimensional, coherently-detected absorption spectrum. The
phase coherent spectra reject signal contributions from long-timescale effects,
as the spectral response is recorded as a function of time delays between pulses
which vary between $100\si{\femto\second}$ and $1\si{\nano\second}$.

For this sample, the linear TPC spectra can be difficult to interpret on their
own, as there are many observed peaks to identify. Thus, we also employ a more
powerful third-order nonlinear spectroscopic technique, multidimensional
coherent spectroscopy (MDCS). Our collinear MDCS variant
\cite{Tekavec2007,Nardin2013}, depicted in Fig. \ref{fig:MDCS} (b), uses a
nested Mach-Zhender interferometer to separate four optical pulses from the Ti:Sapph laser. For the spectra used here, the first ($\tau$) and third ($t$) time
delays are varied, and  these data are Fourier transformed with respect to these
time delays to obtain a two dimensional map of the third-order signal as a
function of excitation ($\omega_\tau$) and detection ($\omega_t$) frequency axes
\cite{Tekavec2007,Nardin2013}. Note that because $\omega_t$ has opposite phase
evolution from $\omega_\tau$, the vertical excitation axis is negative, thus the
diagonal at which the excitation and detection frequencies have equal magnitude
is the downward sloping diagonal. 

There are two main advantages to using multidimensional spectra. First,
we can easily distinguish between homogeneous and inhomogeneous broadening, as
these will broaden the MDCS peak along perpendicular diagonal directions. By
measuring the widths of the peaks along and across the equal frequency diagonal,
we can measure the both the homogeneous and inhomogeneous linewidths
respectively. More important to this work, MDCS spectra can be used to identify
if two spectral peaks are coupled. If coupling exists between two transitions, a
coupling crosspeak will occur where the excitation and detection frequencies
correspond to those of the two coupled transitions. Since a single \ce{SiV-}
center can emit at four different frequencies, a spectrum with several shifted
families of \ce{SiV-} centers can be difficult to interpret. Using a single
multidimensional spectrum, we can determine which spectral peaks correspond to a
single family.

While the information contained in a TPC measurement is not as
rich as that in a full MDCS spectrum, the main advantage is that the acquisition
time for a TPC spectrum is much shorter. This enables us to take a large number
of linear spectra to observe position dependent trends, while only taking a few
select MDCS spectra when needed to assign the peaks.

As mentioned previously, both the TPC and MDCS spectra used here utilize
collinear geometries. This enables a smaller spot size than $k$-vector selection
MDCS experiments\cite{Nardin2013}, and allows us to compare spectra
taken at nearby locations on the sample. Studies of color centers often take
measurements of a single center, with no guarantees that the center is
representative of neighboring centers. In contrast, the high density of our
sample and the ability of MDCS to untangle complicated spectra enable us to
accurately measure the ensemble averaged optical response.

For both of these spectroscopic techniques, we linearly polarized the light
incident on our sample. This was mainly so that the detected light could be
cross-polarized to reduce laser scatter. Additionally, since the four electronic
transitions of the \ce{SiV-} center are polarization
dependent\cite{Rogers2014}, spectra taken at each polarization can be compared
to understand the geometry of the sample. 

\subsection{\label{sec:sample}Sample Information}

Our sample is a chemical-vapor deposition grown, $[110]$-oriented mono
crystalline diamond. An ensemble of  \ce{SiV-} centers was created by
implanting Silicon-29 ions with a focused ion beam, at a depth of
$0.5-2.4\si{\micro\m}$ and number density
$\num{7.5e18}\si{\centi\meter^{-3}}$. The sample was then annealed at
$1000\text{--}1050\si{\celsius}$ and tri-acid cleaned. A picture of the sample
is included in Fig. \ref{fig:bases} (a). The sample has been cleaved, which
explains its unusual shape.

\ce{SiV-} centers can occur along four different orientations in diamond,
corresponding to the four directions of carbon-carbon bonds in the diamond
lattice. Figure \ref{fig:families} (b) shows the four orientations of \ce{SiV-}
centers in our $[110]$-oriented sample, where the four vectors on the diamond
correspond to the \ce{SiV-} axis shown in Fig. \ref{fig:families} (a). Note that we
can group these peaks into in-plane (orange) and out-of-plane (purple)
orientation families.

Figure \ref{fig:families} (c) depicts the level structure of the \ce{SiV-}
zero-phonon line (ZPL). The ZPL has mean frequency $\Delta
_{\text{ZPL}}=407\si{\tera\hertz}$, and the frequencies of the four optical transitions
are also determined by the ground state and excited state splittings,
$\Delta_{\text{gs}}$ and $\Delta_{\text{es}}$. It has been previously
shown that two of the optical transitions are excited by light polarized
perpendicular to the the \ce{SiV-} axis shown in Fig. \ref{fig:families} (a),
and and the other two are excited by light polarized parallel to this
axis\cite{Rogers2014}.

The sample was placed in a closed-loop cryostat at a temperature of
$11\si{\kelvin}$, and the optical experiments were conducted using a home-built
confocal microscope to focus $736\si{\nano\meter}$ pulses from a
Titanium:Sapphire (Ti:Sapph) oscillator onto the sample face. As shown in
Fig. \ref{fig:MDCS}, the sample is tilted by $30^\circ$ from normal to reject
the reflected Ti:Sapph beam and any coherent scatter that could corrupt the PL
measurements. The two linear polarizations of light used here are depicted in
Fig \ref{fig:families} (b).

\section{\label{sec:results}Results and Analysis}

\subsection{\label{sec:identification}Peak Identification}

\begin{figure*}\centering\includegraphics{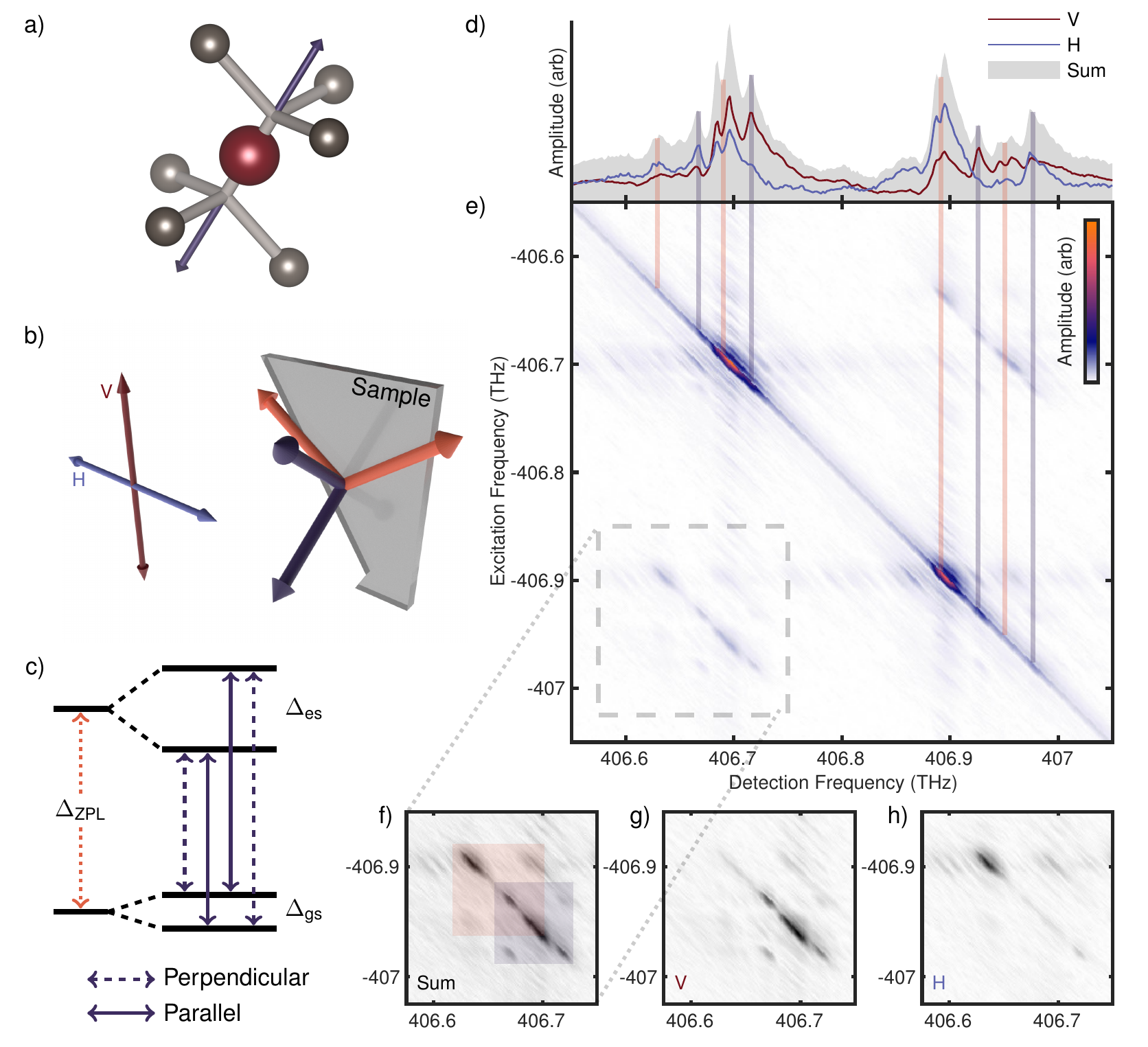}
\caption{\label{fig:families} (a) A depiction of the \ce{SiV-} color
  center\cite{VESTA}, with the red sphere representing a \ce{^29Si} interstitial
  atom between the two lattice vacancies, and the grey spheres representing
  nearby carbon atoms. The \ce{SiV-} axis is shown in
  purple. (b) The four orientations of \ce{SiV-} in our $[110]$ oriented
  diamond. The orientations of the centers along the $\langle111\rangle$ axes of
  the lattice yield two in-plane (orange) and two out-of-plane (purple)
  orientations. The directions of vertical and horizontal polarizations relative
  to the sample are shown. (c) The \ce{SiV-} ZPL level structure, as well as
  the microscopic polarization selection rules with respect to the \ce{SiV-}
  axis. (d) TPC spectra of the \ce{SiV-} ZPL of vertical and horizontal
  polarizations, and their sum of the incident laser from a single point on the
  sample. (e) A rephasing, MDCS spectrum of the \ce{SiV-} ZPL, taken as the sum
  of spectra using both vertical and horizontal polarizations. The orange and
  purple vertical lines highlight the peaks corresponding to the two orientation
  families. (f-g) More detail of the outlined section in the MDCS spectrum,
  including both vertical and horizontal components. In (f), the crosspeak
  families are highlighted.}
\end{figure*}

Previous experimental data taken of \ce{SiV-} centers
\cite{Clark1995,Rogers2014,Meesala2018,Smallwood2020} shows the four
spectral peaks corresponding to the four optical transitions in
Fig. \ref{fig:families} (c). Incoherent photoluminescence spectra taken from our
sample has this structure as well\cite{Smallwood2020}. However, the TPC
photoluminescence spectrum in Fig. \ref{fig:families} (d) shows significantly
more peaks, indicating the presence of multiple groups of color centers.

The TPC measurement contains no information on coherent coupling between
states. Additionally, due to the large strain susceptibility of the system,
small variations in strain could change the ground state splittings on the order
of $10\si{\giga\hertz}$\cite{Meesala2018}, making it difficult to tell which
transitions belong to the same groups of color centers.

MDCS can be used to identify the groups of peaks in the spectra. Spectra taken
with horizontally and vertically incident light were summed to create the
spectrum in Fig. \ref{fig:families} (e). The insets (f-h) below this spectrum
show more detail for some of the crosspeaks, both for the two linear spectra and
the combined spectrum. Note that in the combined spectrum, the seven visible
crosspeaks can be grouped into two squares, which are highlighted in (f). Each
square is formed from two lower energy crosspeaks, and two higher energy
crosspeaks, and the peaks used in each square are distinct.
This allows us to group the spectral peaks into two families, where peaks in a
single family are coupled, and no coupling is observed between different
families. Using this, we conclude that each family of peaks corresponds to a
different type of \ce{SiV-} center. Additionally, the MDCS spectrum shows some
inhomogeneity, which could be due to microscopic strain fluctuations
\cite{Smallwood2020} or interactions between nearby \ce{SiV-} centers
\cite{Day2021}.

To identify the origin of the two families of \ce{SiV-} centers, we can use the
polarization data. While the MDCS spectra can be used for this, it is easier
to compare linear spectra, such as the TPC spectra shown in
Fig. \ref{fig:families} (d). Note that the relative peak strengths of the two
families of peaks under different polarizations of incident light are very
different. This suggests the two families of \ce{SiV-} centers are oriented
differently in the diamond lattice. By relating the different possible
orientations of \ce{SiV-} centers in our 
sample to the polarization selection rules shown in
Fig. \ref{fig:families} (c), we find that the two peak families identified above
correspond to the in-plane and out-of-plane orientations in Fig
\ref{fig:families} (b). While it may be feasible to group the peaks into
families based solely on polarization data, this would become increasingly
difficult for higher amounts of strain, and the crosspeaks in the MDCS spectrum
are easier to interpret and give a higher degree of certainty.

The peaks corresponding to these families are indicated in
Fig. \ref{fig:families} (d-e) using orange and purple vertical lines,
corresponding to the in-plane and out-of-plane families. In the scans seen here,
we do see additional splitting of the in-plane peaks. The specific in-plane
orientation can be determined with additional linearly polarized spectra.

\subsection{\label{sec:strain}Strain Calculation}

\begin{figure*}[]\centering\includegraphics{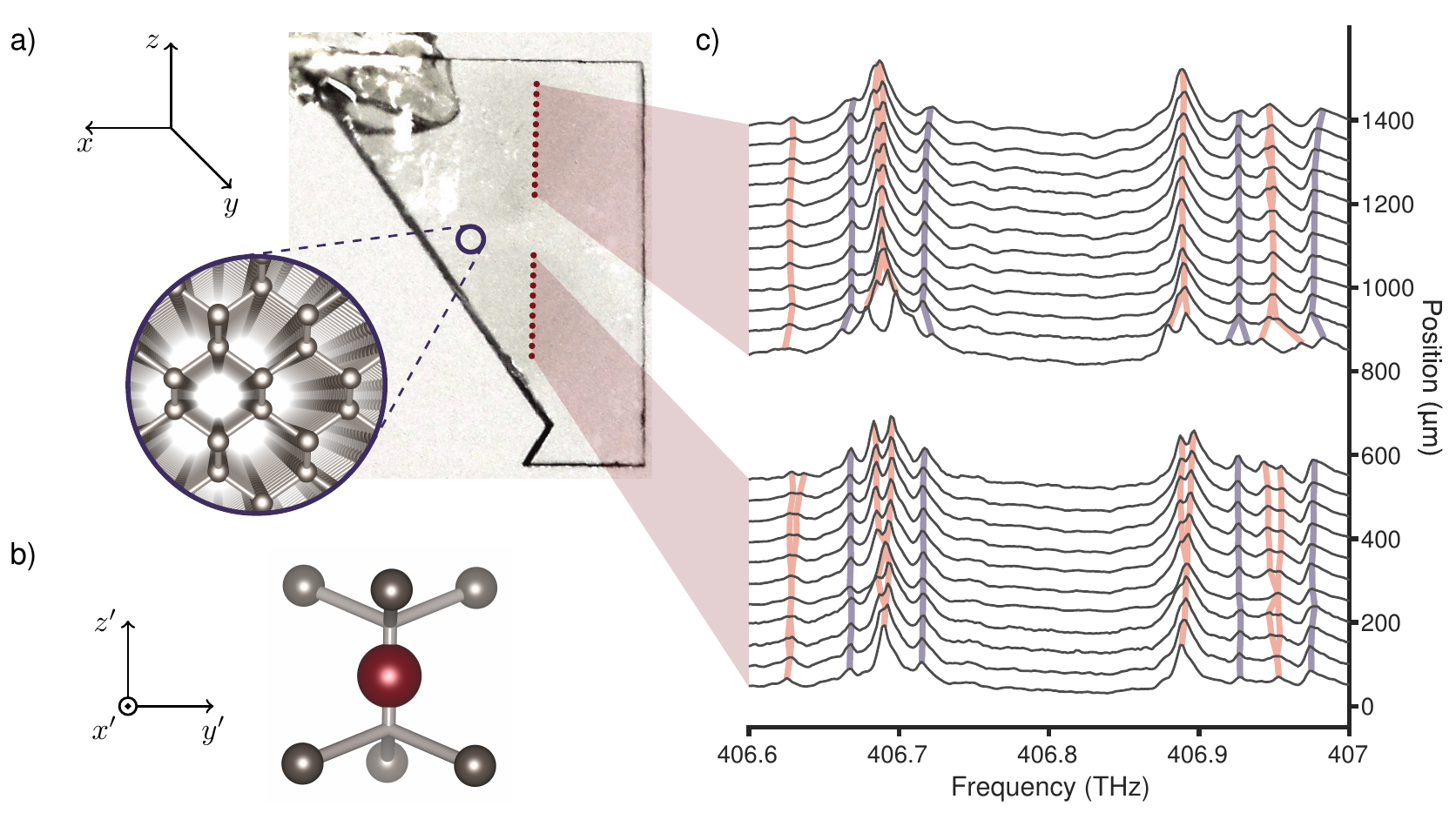}
\caption{\label{fig:bases}(a) An image of the sample illustrating both the
  locations of the $23$ linear spectra used in the strain calculation, as well as
  the crystal basis used for the global strain tensor. (b) The basis of
  a single \ce{SiV-} center\cite{VESTA}. (c) A series of TPC spectra, summing
  both horizontal and vertical polarizations. The centers of each peak are
  highlighted.}
\end{figure*}

While Section \ref{sec:identification} partially explains our spectra, it does
not explain the origin of the peak shifts. We propose that the source of these
shifts is due to strain intrinsic to our sample, mainly because the shifts are
different for different \ce{SiV-} orientations.
By assuming this hypothesis is true, we can solve for the full strain tensor in
our sample. 

Previous work completed by Meesala et al.\cite{Meesala2018} derived equations
relating four strain susceptibility parameters to the six strain tensor indices
local to a given \ce{SiV-} center. They combined these equations with experimental
data and simulated strain tensor data to fit for the strain susceptibility
parameters. In this work, we instead used experimental data and the strain
susceptibility parameters to estimate the strain tensor.

The equations from Ref. \onlinecite{Meesala2018} are
\begin{widetext}
\begin{subequations}\label{eqn:Meesala}
\begin{eqnarray}
\Delta_{\text{ZPL}}
=\Delta_{\text{ZPL,0}}
+\left(t_{\parallel,\text{es}}-t_{\parallel,\text{gs}}\right)\epsilon_{zz}'
+\left(t_{\perp,\text{es}}-t_{\perp,\text{gs}}\right)
\left(\epsilon_{xx}'-\epsilon_{yy}'\right),\label{eqn:ZPL}\\
\Delta_{\text{gs}}
=\sqrt{\lambda_{\text{SO},\text{gs}}
  +4\left(d_{\text{gs}}\left(\epsilon_{xx}'-\epsilon_{yy}'\right)
    +f_{\text{gs}}\epsilon_{zx}'\right)
  +4\left(2d_{\text{gs}}\epsilon_{xy}'
  +f_{\text{gs}}\epsilon_{yz}'\right)^2},\label{eqn:gs}\\
\Delta_{\text{es}}
=\sqrt{\lambda_{\text{SO},\text{es}}
  +4\left(d_{\text{es}}\left(\epsilon_{xx}'-\epsilon_{yy}'\right)
    +f_{\text{es}}\epsilon_{zx}'\right)
  +4\left(2d_{\text{es}}\epsilon_{xy}'
  +f_{\text{es}}\epsilon_{yz}'\right)^2}.\label{eqn:es}
\end{eqnarray}
\end{subequations}
\end{widetext}
Note that Eqs. \ref{eqn:gs} and \ref{eqn:es} are linearly dependent, as the ground
state and excited state splitting respond similarly to strain. These equations
are given in the \ce{SiV-} reference frame, which is depicted in
Fig. \ref{fig:bases} (b). Since there are four different orientations of \ce{SiV-}
centers in four distinct strain environments, we must first rotate these
equations into a common basis. We choose the crystal basis depicted in Fig
\ref{fig:bases} (a). For instance, for one of the in-plane peaks we have
\begin{subequations}
  \begin{eqnarray}
    \epsilon_{xx}'
    =&\tfrac13\epsilon_{xx}-\tfrac{2\sqrt2}3\epsilon_{zx}+\tfrac23\epsilon_{zz},\\
    \epsilon_{yy}'=&\epsilon_{yy},\\
    \epsilon_{zz}'
    =&\tfrac23\epsilon_{xx}+\tfrac{2\sqrt2}3\epsilon_{zx}+\tfrac13\epsilon_{zz},\\
    \epsilon_{xy}'
    =&\tfrac{\sqrt3}3\epsilon_{xy}-\tfrac{\sqrt6}3\epsilon_{yz},\\
    \epsilon_{yz}'
    =&\tfrac{\sqrt6}3\epsilon_{xy}+\tfrac{\sqrt3}3\epsilon_{yz},\\
    \epsilon_{zx}'
    =&\tfrac{\sqrt2}3\epsilon_{xx}-\tfrac{\sqrt2}3\epsilon_{zz}-\tfrac13\epsilon_{zx}.
  \end{eqnarray}
\end{subequations}
This calculation is repeated for the three other orientations, and the results
are substituted into Eqs. \ref{eqn:Meesala}, yielding twelve total equations,
eight of which are independent. Since (in the crystal basis) there are six
strain tensor indices, we can use these equations to solve for the full strain
tensor.

We took a series of TPC spectra at $23$ locations on our sample, as shown in
Fig. \ref{fig:bases} (c). These locations lie along a single line, with a gap in the
middle where the implantation density is lower and reliable peak locations and
strain estimates could not be measured. Slight shifts in the frequencies of the
spectral peaks can be seen.

To use this data to solve for the strain, we first uses the techniques in Section
\ref{sec:identification} to identify the peaks in our linear spectra (several
MDCS spectra were taken at select locations to help with this). The peaks were
fitted to find the locations of all $16$ spectral peaks for each scan, taking
advantage of both vertical and horizontal polarization spectra, since some peaks are
more visible on a given polarization. In cases where two peaks could not be
resolved (which sometimes occurs for two in-plane or two out-of-plane peaks),
a single frequency was reported.

Next, we used Eqs. \ref{eqn:Meesala} to find analytical expressions for the
frequencies of the $16$ spectral peaks as a function of the six strain tensor
indices in the crystal basis, $p_i^{\text{calc}}\left(\varepsilon\right)$. We
estimated the error in our measured peaks $p_i^{\text{meas}}$ to be
$\sigma=2\si{\giga\hertz}$. This gives us an expression for $\chi^2$,
\begin{equation}\label{eqn:chi2}
  \chi^2
  =\sum_{i=1}^{16}\frac{\left(p_i^{\text{meas}}-p_i^{\text{calc}}\left(\varepsilon\right)\right)^2}{\sigma^2}.
\end{equation}
Next we find the strain tensor $\varepsilon$, or equivalently the six strain
tensor indices, such that $\chi^2$ is minimized. This tensor is our solved
result.

To estimate the error in the calculation we note that if the error in our peak
locations is random, then the $\chi^2$ value from Eqn. \ref{eqn:chi2} should
indeed follow a chi-square distribution with $16$ degrees of freedom for the
true value of the strain tensor. Thus, there is a $90\%$ chance that
$\chi^2<23.5$. We explored the parameter space local to our solved strain
tensor to find the range of strain tensors such that $\chi^2$ is less than the
desired value.

The strain solving algorithm was also tested on simulated data. Peaks were
generated with Eqs. \ref{eqn:Meesala} using random values for the strain tensor
indices. Random error was added to these peak values, and we attempted to
recover the original strain tensor indices using the algorithm above. The
algorithm worked as well as expected within the range of observed strain
values. The algorithm begins to break down as the sheer strain tensor index
$\varepsilon_{xy}$ becomes large (greater than $\num{0.8e-5}$).

\subsection{\label{sec:results}Strain Results}

\begin{figure}[h!]\centering\includegraphics{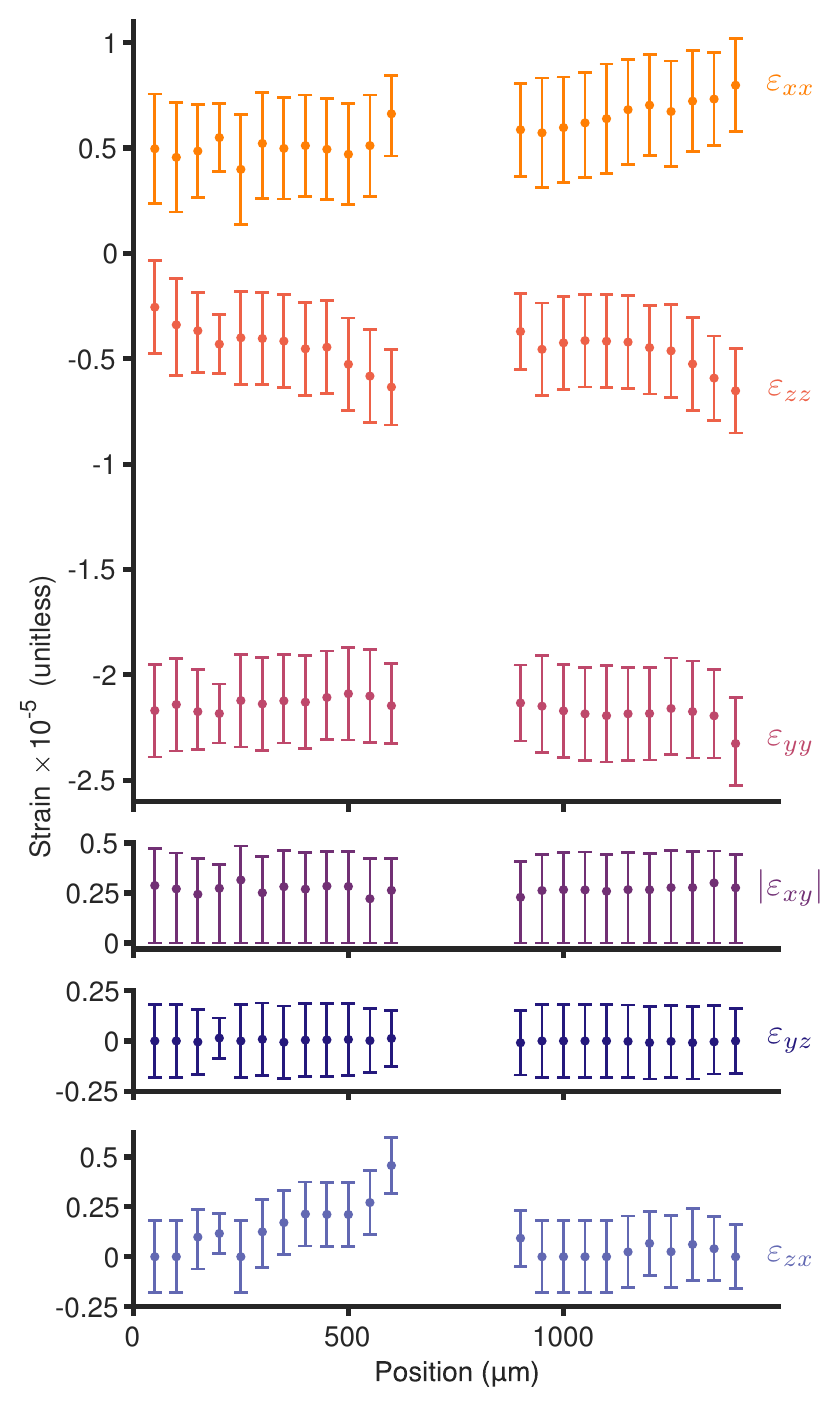}
\caption{\label{fig:strain}Calculated strain tensor indices across the points
  shown in Fig. \ref{fig:bases}. The first plot shows the normal strain tensor
  indices, $\varepsilon_{xx}$, $\varepsilon_{yy}$, and $\varepsilon_{zz}$. The
  following three plots show the sheer strain tensor indices,
  $\varepsilon_{xy}$ (for which positive values are plotted due to a sign
  ambiguity), $\varepsilon_{yz}$ and $\varepsilon_{zx}$. $90\%$ confidence
  intervals are shown.}
\end{figure}

The results are shown in Fig. \ref{fig:strain}. Before drawing conclusions from
these results, some limitations with the data should be addressed. First, the
values of the normal strain tensor indices are dependent on the mean ZPL
frequency at zero strain, or $\Delta_{\text{ZPL},0}$. We did not take a
direct measurement of this, but we estimated it to be
$406.795\si{\tera\hertz}$. However, we find that changing this value results in a
constant offset of the normal strain tensor indices $\varepsilon_{xx}$,
$\varepsilon_{yy}$ and $\varepsilon_{zz}$, and their relative values are
preserved. Next, a sign ambiguity for small values
of $\epsilon_{xy}$ means we cannot find the sign of this parameter, so the plot
in Fig. \ref{fig:strain} uses positive $\epsilon_{xy}$. For all
indices, the error bars in Fig. \ref{fig:strain} are only meant to represent
error due to random error in the peak locations. They do not account for
systematic errors, such as errors in the strain susceptibility parameters used
in Eqs.\ref{eqn:Meesala}.

Despite these limitations, we do see a varying strain in our sample. The normal
strain indices are nonzero, and these values appear to vary across the
sample. While we do not see statistically significant magnitudes for the sheer
strain tensor indices $\varepsilon_{xy}$ and $\varepsilon_{yz}$, the sheer
strain tensor index $\varepsilon_{zx}$ does show statistically significant
variation. 

Previous strain measurements \cite{Meesala2018} observe shifts in single
\ce{SiV-} centers. By contrast, these measurements are taken on an ensemble of
\ce{SiV-} centers, as mentioned in Section \ref{sec:spectroscopy}. This means
we measure the strain tensor averaged over the laser spot, rather than at
specific centers. Note that this variation in macroscopic strain is also
different from the microscopic strain fluctuations proposed in
Ref. \onlinecite{Smallwood2020}, as the latter concerns individual \ce{SiV-}
centers in highly strained environments.

The source of this strain is likely due to the implantation and annealing
processes. As mentioned in Section \ref{sec:sample}, our sample has a rather
high number density of implanted silicon of
$\num{7.5e18}\si{\centi\meter^{-3}}$. This density corresponds to about one
silicon atom for every $\num{2.3e4}$ carbon atoms. If we assume, as an
approximation, implanting the silicon does not increase the size of the sample,
but merely pushes the atoms uniformly closer together, then we can estimate the
normal strain as $-\num{1.4e-5}$. While this simple estimation does not take
into account many complexities of the system, it does agree with our
experimental values to within an order of magnitude.

\section{\label{sec:conclusion}Conclusion}

By analyzing both MDCS and TPC measurements of \ce{SiV-} centers, we were able
to measure the full strain tensor of our sample. We do measure a nonzero strain,
which varies across the sample. This strain is most likely due to the large
amount of implanted silicon in our sample. In future work, we hope to take
measurements on a variable density sample, to further measure the relationship
between implantation density and strain. This work may also be useful in using
\ce{SiV-} centers in diamond as a high pressure strain gauge, potentially in a
diamond anvil cell.

\section*{Acknowledgements}

T. S. acknowledges support from the Federal Ministry of Education and Research of Germany (BMBF, project DiNOQuant, 13N14921). Ion implantation work to generate the SiV\textsuperscript{--} centers was performed, in part, at the Center for Integrated Nanotechnologies, an Office of Science User Facility operated for the U.S. Department of Energy (DOE) Office of Science. Sandia National Laboratories is a multi-mission laboratory managed and operated by National Technology and Engineering Solutions of Sandia, LLC., a wholly owned subsidiary of Honeywell International, Inc., for the DOE's National Nuclear Security Administration under contract DE-NA-0003525. This paper describes objective technical results and analysis. Any subjective views or opinions that might be expressed in the paper do not necessarily represent the views of the DOE or the United States Government.

\bibliography{strain}

\end{document}